\documentclass[prb,twocolumn,showpacs,preprintnumbers,superscriptaddress]{revtex4-1}
\usepackage[usenames,dvipsnames]{color}
\usepackage{amssymb} 
\usepackage{times}
\usepackage{graphicx}
\usepackage{graphicx}
\usepackage{dcolumn}
\usepackage{datetime}
\usepackage{soul}
\usepackage{subfigure}
\usepackage[bookmarks, colorlinks=true, plainpages = false,citecolor = red, urlcolor = black, 
filecolor = blue]{hyperref}
\usepackage{multirow} 
\usepackage[english]{babel}
\usepackage{graphicx}
\usepackage{amsmath}
 
\usepackage[english]{babel}
\usepackage{graphicx}
\usepackage{amsmath}
 
\begin{document}
\title{Nearly perfect spin-filtering in curved two-dimensional topological insulators}
 
\author{Alireza Saffarzadeh} 
\affiliation{Department of Physics, Simon Fraser
University, Burnaby, British Columbia, Canada V5A 1S6}
\affiliation{Department of Physics, Payame Noor University, P.O.
Box 19395-3697 Tehran, Iran}

\author{George Kirczenow} 
\affiliation{Department of Physics, Simon Fraser
University, Burnaby, British Columbia, Canada V5A 1S6}

\date{\today}

\begin{abstract}\noindent

The spintronic properties of curved nanostructures derived from two-dimensional topological insulators (2DTI's) are explored theoretically
with density functional theory-based (DFT) calculations and tight-binding models. We show that curved geometries make
it possible to manipulate electron spins in ways that are not available for planar 2DTI devices.
We predict that, unlike planar 2DTI devices, curved  2DTI-related
nanostructures can function as highly effective {\em two}-terminal spin filters even in the absence of magnetic fields.
We construct a generalization to curved geometries 
of our previous tight binding model of the wide band gap planar 2DTI bismuthene on SiC. The resulting model, applied to an 
ideal dome geometry with a free edge, is shown to exhibit quantum spin Hall physics, including spin polarized edge states.
The model predicts nearly perfect spin filtering by the dome for a particular two-terminal geometry in the absence 
of magnetic fields. Our DFT calculations predict a Bi$_{105}$Si$_{105}$H$_{15}$ dome of bismuthene with
adsorbed silicon and hydrogen atoms to be stable. Our tight binding model, adjusted to match density of states given
by DFT calculations, predicts that the
Bi$_{105}$Si$_{105}$H$_{15}$ dome should exhibit quantum spin Hall physics and very effective spin filtering in a two-terminal 
arrangement. 

\end{abstract}

 \maketitle
 
 \section{Introduction}
\label{Intro}

In two-dimensional topological insulators (2DTIs) electron transport occurs via edge states when the Fermi level is located in the bulk band gap.\cite{KM1,KM2,Bernevig2006,FuKane2007,Moore2007,Konig2007,Hasan2010,Qi2011,Ando2013,TIcourse}
These edge states exhibit locking between the electron momentum and spin which results in a quantum spin Hall (QSH) effect,\cite{KM1,KM2,Bernevig2006,FuKane2007,Moore2007,Konig2007} i.e., the electric current is carried by spin polarized electrons with opposite spin orientations at the opposite edges of the sample. The 2D QSH systems considered to date have been planar; 
\cite{Hasan2010,Qi2011,Ando2013,TIcourse} curved 2D QSH systems have received little if any attention theoretically or experimentally. In this paper we initiate the exploration of the physics of curved 2D QSH systems theoretically by constructing tight binding models of finite  nanostructures based on 2DTIs with curved geometries and examining their properties. We find that curved geometries offer additional degrees of freedom for manipulating the orientations of the spins of  QSH edge states that are not available for planar 2DTIs. In particular, while the spin orientation of an electron traveling along the edge of a planar 2DTI nanostructure remains fixed, we show that the spin orientation of an electron traveling along the edge of a suitably curved 2DTI can vary as a function of the position along the edge. Thus while the spin orientations induced by an electric current on the opposite edges of planar 2DTIs are always antiparallel, our work reveals that, for suitable curved geometries, the spin orientations on opposite edges of the 2DTI can be tailored so as to vary continuously from antiparallel to parallel depending on the locations where the spin orientations are measured. From the perspective of potential applications, it follows that appropriately contacted curved 2DTIs can function as 2-terminal spin filters in the absence of magnetic fields whereas in order to achieve spin filtering at zero magnetic field in a planar 2DTI at least a 3-terminal device is required.

 \section{Curved Bismuthene Nanostructures}
\label{Binano}

Recent theoretical and experimental work has provided strong evidence that monolayer bismuthene on SiC is a wide gap 2DTI.\cite{Hsu2015,Reis2017,Dominguez2018,GLi2018,GK2018,Canonico2019,Azari2019,Hao2019,Stuhler2020} Tight binding models of the bismuthene monolayer employing basis sets consisting {\em only} of the valence orbitals of the bismuth atoms but parameterized so as to take into account the influence of the SiC substrate on the bismuthene have been constructed.\cite{Reis2017,Dominguez2018,GLi2018,GK2018,Canonico2019,Azari2019,Hao2019} These simple models have succeeded in capturing the essential topological insulator and QSH physics of this planar system.\cite{Reis2017,Dominguez2018,GLi2018,GK2018,Canonico2019,Azari2019,Hao2019} 

Whether {\em curved} bismuthene monolayers chemically modified by suitable adsorbates can also exhibit topological insulator and QSH physics is unknown at present. However, theoretical studies have suggested that approximately spherical and cylindrical bismuthene nanostructures may be stable.\cite{Su2002,Zdetsis2010,Kharissova2012,Jin2017} Also, bismuth nanotubes have been synthesized.\cite{Li2001,Yang2003,Li2006,Yang2008,Ma2010,Derrouiche2010a,Derrouiche2010b,Boldt2010}

\begin{figure}[t]
\centering
\includegraphics[width=0.7\linewidth]{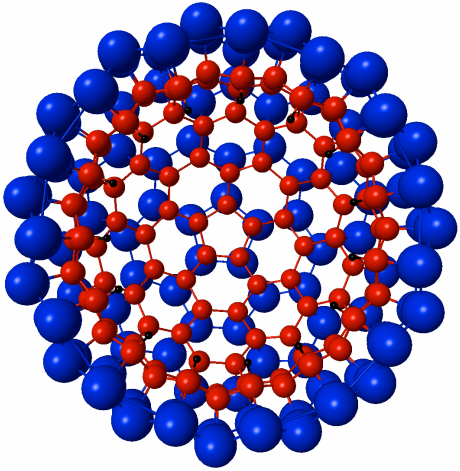}
\caption{(Color online).
Relaxed Bi$_{105}$Si$_{105}$H$_{15}$ dome. 105 bismuth atom (blue) bismuthene dome with zigzag edge stabilized by 105 silicon 
(red) atoms viewed from the concave side. Edge Si atoms are passivated with H atoms (black). Image prepared using Macmolplt software.\cite{MacMolPlt}}
\label{hemi} 
\end{figure}

We have investigated curved bismuthene-based nanostructures further by means of density functional theory (DFT) computer simulations as implemented in the GAUSSIAN 16 package with the B3PW91 functional and Lanl2DZ effective core potential and basis sets \cite{Frisch}.  The electronic energy and ionic forces of our optimized geometries were converged within 10$^{-5}$ eV and 0.0008 eV/\AA, respectively.
Our simulations indicate that an approximately spherical bismuthene fullerene structure consisting of 180 Bi atoms comprised of Bi hexagons 
and 12 symmetrically arranged Bi pentagons should be stable and that a similar 180 Bi atom fullerene with 180 Si atoms in its interior, one Si atom bound to each Bi atom, should should also be stable. We attribute the stability of the latter structure to the fact that for it the distance between nearest neighbor Si atoms is similar to the nearest neighbor distance in crystalline silicon; we find that bismuthene fullerenes decorated with Si atoms but having very different nearest neighbor Si-Si distances tend to be unstable. We also find that spherical domes of bismuthene are unstable, being prone to collapse into compact clusters of Bi 
atoms. However, interestingly, we find that a particular bismuthene dome can be stabilized by the addition of Si and H atoms. In Fig.\ref{hemi} we show such a stable structure of 105 Bi and 105 Si atoms obtained by relaxing a truncated 180 Bi-Si atom spherical fullerene. A Si atom is bound to each Bi atom on the concave surface of the dome; the dome has a zigzag edge. The edge Si atoms are passivated with hydrogen atoms. We refer to this nanostructure as a Bi$_{105}$Si$_{105}$H$_{15}$ dome.  This suggests that making a curved monolayer bismuthene nanostructure with a free edge may be possible.
  
Whether such stable curved nanostructures based on monolayer bismuthene or other 2DTI's and having a free edge can be realized experimentally remains an open question. Never the less it is of interest to explore the properties of such prospective systems theoretically. Here we initiate this theoretical work by constructing and studying potentially relevant tight binding models. We construct a generalization of our tight-binding Hamiltonian of planar bismuthene (modified chemically by SiC)\cite{GK2018} to curved bismuthene nanostructures. We show that for an {\em ideal} 105 Bi atom spherical dome geometry this basic model exhibits spin polarized edge states, a QSH effect and nearly perfect spin filtering in a two terminal arrangement. 

We then modify this basic model in such a way as to reproduce approximately the partial density of states on the Bi atoms obtained by our DFT calculations for the {\em relaxed} Bi$_{105}$Si$_{105}$H$_{15}$ dome shown in Fig. \ref{hemi}. We find that the electronic structure and transport properties of this realistic curved bismuthene-based nanostructure are more complex, being strongly affected by both edge and bulk states, but that it also exhibits a pronounced QSH effect and very effective spin filtering in a two terminal arrangement.
  
\section{Generalization of the tight binding model of infinite monolayer bismuthene on SiC to curved geometries}
\label{Model}

The bismuth atoms of planar monolayer bismuthene on SiC form a honeycomb lattice.\cite{Hsu2015,Reis2017} For bismuthene in the $x$-$y$ plane,  the main contributions to the low energy electronic states are those of the Bi $6p_x, 6p_y$ and $6s$ atomic valence orbitals.\cite{Reis2017} The tight binding models of bismuthene on SiC have to date have employed basis sets consisting of only these bismuth valence orbitals. In bismuthene on SiC the Bi $6p_z$ valence orbital is shifted away from the Fermi level because of the interaction with the SiC substrate.\cite{Reis2017} For this reason it has been omitted in most tight binding models of planar bismuthene on SiC. However, for curved bismuthene geometries the Bi $6p_z$ orbital should be included in the theory. We do this as follows. 

Our generalized tight binding model Hamiltonian is of the form
\begin{equation}
\begin{split}
H_{i \alpha s,i' \alpha' s'}&=H^{i}_{\alpha,\alpha'}\delta_{i,i'}\delta_{s,s'}+H^{\text{NN}}_{i \alpha,i' \alpha'}\delta_{s,s'}+\\&+H^{\text{SO}}_{\alpha s,\alpha' s'}\delta_{i,i'}+H^{\text{R}}_{\alpha s,\alpha' s'}\delta_{i,i'}
\label{Ham}
\end{split}
\end{equation}
where $\alpha~\text{and}~\alpha'$ denote the Bi $6p_x, 6p_y,  6p_z$ and $6s$ valence orbitals of atoms $i~\text{and}~i'$, $s~\text{and}~s'$ are
spin indices, $H^{i}_{\alpha,\alpha'}$ is the onsite orbital part of the Hamiltonian matrix for atom $i$ omitting the spin-orbit and Rashba contributions, $H^{\text{NN}}_{i \alpha,i' \alpha'}$ is the Hamiltonian matrix element between orbital $\alpha'$ on Bi atom $i'$ and orbital $\alpha$ on Bi atom $i$ that is a nearest neighbor of atom $i'$, $H^{\text{SO}}$ is the atomic spin-orbit interaction and $H^{\text{R}}$ is the Rashba Hamiltonian. Following the reasoning in Ref. \onlinecite{Reis2017} only the intra-atomic matrix elements of $H^{\text{SO}}$ and $H^{\text{R}}$ are considered here.

The onsite orbital Hamiltonian matrix elements $H^{i}_{\alpha,\alpha'}$ are given in Table \ref{onsite}. $H^{i}_{\alpha,\alpha'}$ has energy eigenvalues
$E_s$ and $E_r$ corresponding to the orbital states  $|6s\rangle$ and $|6p_r\rangle = a|6p_x\rangle+b|6p_y\rangle+c|6p_z\rangle$, respectively, where $\hat{r}=(a,b,c)$ is the unit vector normal to the surface of the curved topological insulator at Bi atom $i$. The other two eigenvalues of $H^{i}_{\alpha,\alpha'}$  correspond to the $6p$ states that are orthogonal to $|6p_r\rangle$ and are both zero. Thus $|6p_r\rangle$, the $6p$ orbital whose symmetry axis is parallel to the local normal to the surface of  curved topological insulator, is shifted in energy relative to the other $6p$ orbitals by an amount $E_r$, emulating the shift of the $6p_z$ orbital relative to $6p_x$ and $6p_y$ for planar bismuthene on SiC.\cite{Reis2017}

\begin{table}[t]
\caption{The onsite orbital part $H^{i}_{\alpha,\alpha'}$ of the model Hamiltonian matrix for atom $i$  in Eq. \ref {Ham}. 
Here $\hat{r}=(a,b,c)$ is the unit vector normal to the surface of the curved topological insulator at Bi atom $i$. The parameter values are $E_s=-10.22$eV and $E_r=-5.0$eV.}
\begin{center}
\begin{tabular}{l|c|c|c|c}
 $H^{i}_{\alpha,\alpha'}$ & $6s'$ & $6p'_x$ & $6p'_y$& $6p'_z$\\ 
 \hline
 $6s$ & $E_s$ & 0 &  0 &  0\\
  \hline
 $6p_x$ &  0 &$ a^2E_r$ &  $ abE_r$ & $ acE_r$\\
  \hline
 $6p_y$ &  0 & $baE_r $ & $b^2E_r $& $bcE_r $\\
   \hline
 $6p_z$ &  0 & $caE_r $ & $cbE_r $& $c^2E_r $\\
   \hline
 
\end{tabular}
\end{center}
\label{onsite}
\end{table}

The nearest neighbor Hamiltonian matrix elements $H^{\text{NN}}_{i \alpha,i' \alpha'}$  are given in Table \ref{hopping}. For $\delta=0$ they have been fitted to the
band structure\cite{Reis2017} of planar bismuthene on SiC. For geometries other than that of planar bismuthene on SiC, they are assumed to depend on the Bi-Bi bond orientations as in the Slater-Koster 
model\cite{SK1954} and to scale with the bond lengths as in extended H\"{u}ckel theory.\cite{Ammeter}

\begin{table*}[h]
\caption{Nearest neighbor Hamiltonian matrix elements $H^{\text{NN}}_{i \alpha,i' \alpha'}$  in Eq. \ref {Ham}. 
In terms of the coordinates of the primed and unprimed neighboring atoms $d= ((x'-x)^2 + (y'-y)^2+ (z'-z)^2)^\frac{1}{2}; l=(x'-x)/d; m=(y'-y)/d; n=(z'-z)/d$.
Fitting parameter values are $\Sigma= -0.81(1-1.583\delta+0.821\delta^2) \text{eV}, 
\Sigma'= -1.00(1-1.087\delta+0.313\delta^2) \text{eV}, 
\Sigma''= -1.635(1-0.509\delta-0.187\delta^2) \text{eV},
\Pi=0.55(1-1.361\delta+0.634\delta^2)\text{eV}$ and $\delta=d-3.089$\AA.}
\begin{center}
\begin{tabular}{l|c|c|c|c}
 $H^{\text{NN}}_{i \alpha,i' \alpha'}$ & $6s'$ & $6p'_x$ & $6p'_y$& $6p'_z$\\ 
 \hline
 $6s$ & $\Sigma$ & $-\Sigma' l$ &  $-\Sigma' m$ &  $-\Sigma' n$\\
  \hline
 $6p_x$ &  $\Sigma' l$ &$\Sigma'' l^2 +\Pi (1-l^2)$ &  $(\Sigma''-\Pi ) lm$&  $(\Sigma''-\Pi ) ln$\\
  \hline
 $6p_y$ &  $\Sigma' m$ & $(\Sigma''-\Pi ) lm$ & $\Sigma'' m^2 +\Pi (1-m^2)$& $(\Sigma''-\Pi ) mn$\\
   \hline
 $6p_z$ &  $\Sigma' n$ & $(\Sigma''-\Pi ) ln$ & $(\Sigma''-\Pi ) mn$& $\Sigma'' n^2 +\Pi (1-n^2)$\\
   \hline

\end{tabular}
\end{center}
\label{hopping}
\end{table*}

The intra-atomic matrix elements of the spin-orbit Hamiltonian can be approximated as \cite{PRBrapid,PRB}
\begin{equation}
%\nonumber
H^{\text{SO}}_{\alpha s,\alpha' s'} =\zeta_{l} \frac{\langle C_\alpha s | \mathbf S \cdot \mathbf L | C_{\alpha'} {s}' \rangle}{\hbar^2}   
\label{intraSO}
\end{equation}
where $\mathbf S~\text{and}~\mathbf L$ are the spin and orbital angular momentum operators, and $C_\alpha$ is the cubic
harmonic that corresponds to orbital state $\alpha$. $\zeta_l$ is the spin-orbit interaction strength and $l$ is the orbital
angular momentum quantum number.\cite{CS} The matrix
$ \langle C_\alpha s | \mathbf S \cdot \mathbf L | C_{\alpha'} {s}' \rangle/{\hbar^2}$ is given in Table \ref{SO}. $\zeta_l$
is regarded here as a model fitting parameter with value $\zeta_1 = 1.8$eV for the Bi $6p$ valence orbitals.

\begin{table}[b]
\caption{Matrix elements of $\frac{\mathbf S \cdot \mathbf L }{\hbar^2}$ that enter the intra-atomic spin-orbit Hamiltonian matrix, Eq. \ref{intraSO}. All matrix elements involving the atomic $s$ orbital are zero.}
\begin{center}
\begin{tabular}{l|c|c|c|c|c|c}
 $\frac{\langle C_\alpha s | \mathbf S \cdot \mathbf L | C_{\alpha'} {s}' \rangle}{\hbar^2}$ & $6p'_x \uparrow'$ & $6p'_x \downarrow'$ 
 & $6p'_y \uparrow'$&$6p'_y \downarrow'$ & $6p'_z \uparrow'$&$6p'_z \downarrow'$\\ 
 \hline
 $6p_x \uparrow$ & $0$ & $0$ &  $-i/2$&0&  $0$& $1/2$\\
  \hline
 $6p_x \downarrow$ &  $0$ &$0$ &  $0$& $i/2$&  $-1/2$ & $0$\\
  \hline
 $6p_y \uparrow$ &  $i/2$ & $0$ & $0$&0&  $0$& $-i/2$\\
   \hline
  $6p_y \downarrow$ &  $0$ & $-i/2$ & $0$&0& $-i/2$ & $0$\\
   \hline
   $6p_z \uparrow$ &  $0$ & $-1/2$ & $0$&i/2& $0$&0\\
   \hline
  $6p_z \downarrow$ &  $1/2$ & $0$ & $i/2$&0& $0$&0\\
   \hline

\end{tabular}
\end{center}
\label{SO}
\end{table}

Rashba phenomena \cite{BR1,BR2} are due to spin-orbit coupling in systems whose
symmetry is broken by the presence of a surface or interface. The form of the
intra-atomic Rashba Hamiltonian matrix elements $H^{\text{R}}_{\alpha s,\alpha' s'}$
for the present system can be deduced by considering a contribution to $\nabla{V(\bf{r})}$ in the general 
spin-orbit Hamiltonian\cite{Kittel}
$\frac{\hbar}{(2mc)^2}\boldsymbol{\sigma}\cdot  \nabla{V(\bf{r}) \times \mathbf{p}  } $
that points in the direction of the local normal to the curved bismuthene surface. The
resulting matrix elements are given in Table \ref{Rashba}.

\begin{table*}[t]
\caption{Matrix elements of the intra-atomic Rashba Hamiltonian $H^{\text{R}}$, Eq. \ref{Ham}. 
Here $\hat{r}=(a,b,c)$ is the unit vector normal to the surface of the curved topological insulator at Bi atom $i$.The 
fitting parameter value is $R=0.395$eV.}%this is for Si inside the buckyball
\begin{center}
\begin{tabular}{l|c|c|c|c|c|c|c|c}
 $H^{\text{R}}_{\alpha s,\alpha' s'}$& $6s' \uparrow'$&$6s' \downarrow'$ &$6p'_x \uparrow'$ & $6p'_x \downarrow'$ 
 & $6p'_y \uparrow'$&$6p'_y \downarrow'$& $6p'_z \uparrow'$&$6p'_z \downarrow'$\\ 
 \hline
  $6s \uparrow$ &0&0&$-iRb$ & $Rc$ &  $iRa$&$-iRc$&  $0$&$iR(b+ia)$\\
  \hline
 $6s \downarrow$ &0&0&$-Rc$ & $iRb$ &  $-iRc$&$-iRa$&  $iR(b-ia)$&  $0$\\
  \hline
 $6p_x \uparrow$ &$iRb$&$-Rc$&$0$ & $0$ &  $0$&0&  $0$& 0\\
  \hline
 $6p_x \downarrow$ &$Rc$&$-iRb$& $0$ &$0$ &  $0$& 0&  $0$& 0\\
  \hline
 $6p_y \uparrow$ &$-iRa$&$iRc$& $0$ & $0$ & $0$&0&  $0$& 0\\
   \hline
  $6p_y \downarrow$ &$iRc$&$iRa$&  $0$ & $0$ & $0$&0&  $0$& 0\\
   \hline
 $6p_z \uparrow$ &0&$-iR(b+ia)$& $0$ & $0$ & $0$&0&  $0$& 0\\
   \hline
  $6p_z \downarrow$ &$-iR(b-ia)$&0&  $0$ & $0$ & $0$&0&  $0$& 0\\
   \hline

\end{tabular}
\end{center}
\label{Rashba}
\end{table*}

The present model is a generalization of our previous model\cite{GK2018} of the planar large gap topological insulator monolayer bismuthene
on SiC. For a monolayer of bismuth atoms arranged on the planar honeycomb lattice of
bismuthene, this basic tight-binding model provides a good approximation to the low energy electronic structure of the planar topological insulator monolayer bismuthene
on SiC, similar to that given by our previous model.\cite{GK2018} Specifically, it yields a low energy band structure of infinite planar 2D bismuthene on SiC close to that shown in Fig.1 of Ref.\onlinecite{GK2018}, including the conduction band minimum at the $\Gamma$ point, the valence band maximum at $K$, the 0.86 eV indirect band gap, and the 0.46eV Rashba splitting of the valence band maximum. For planar zigzag and armchair bismuthene nanoribbons, this model also yields band structures similar to those in Fig.2 of Ref.\onlinecite{GK2018}, including spin-polarized edge states in the bulk band gap with spin-momentum locking, all as expected for nanoribbons of 2D topological insulators.

\section{Transport}
\label{Transport}

In order to study the spin Hall effect and spin filtering of curved bismuthene-based topological insulators
we carry out electronic transport calculations within the Landauer formalism\cite{Econ81,Fish81} according to which
the two terminal source-drain conductance $G$ of a nanostructure at zero temperature in hte linear response regime is given by
\begin{equation}
G = \frac {e^2}{h} T(E_\text{F})
\label{Landauer}
\end{equation}
where the electron transmission probability through the nanostructure at the Fermi energy is
\begin{equation}\label{T}
T(E_\text{F})=\sum_{\alpha,i,\beta,j}|t_{\beta, j, \alpha, i}(E_\text{F})|^2\frac{v_{\beta,j}}{v_{\alpha,i}}.
\end{equation}
Here $t_{\beta, j, \alpha, i}$ is the amplitude for electron scattering at the Fermi energy from state $\alpha$ of 1D lead $i$
connected to the electron source to state $\beta$ of 1D lead $j$
connected to the electron drain reservoir. $v_{\alpha,i}$ and $v_{\beta,j}$ are the corresponding subband Fermi velocities.

Since in this work we are interested in the spin Hall effect and spin filtering, we consider spin-unpolarized electrons
entering the device through the electron source contact and calculate the spin resolved probabilities $T_{\uparrow}$ and 
$T_{\downarrow}$ of spin up and spin down electrons
exiting through the drain contact at the Fermi energy. $T_{\uparrow}$ and 
$T_{\downarrow}$ are obtained by restricting the sum over $\beta$ in Eq. (\ref{T}) to spin up and spin down states,
respectively, while including both the spin up and spin down states in the sum over $\alpha$. The axis of quantization is the $z$-axis.
  
The scattering amplitudes $t_{\beta, j, \alpha, i}$ are obtained by solving numerically the Lippmann-Schwinger equation 

\begin{equation}\label{lippmann}
|\psi\rangle=|\phi_{\circ}^{\alpha,i}\rangle+G_{\circ}(E)V|\psi\rangle.
\end{equation}
Here $|\phi_{\circ}^{\alpha,i}\rangle$ is an eigenstate of the $i^{th}$ ideal 1D lead that is decoupled from the nanostructure consisting of the quantum dot and conducting contacts (if those are present), $G_{\circ}(E)$ is the sum of the Green's functions of the nanostructure and 1D leads if they are decoupled from the nanostructure, and $|\psi\rangle$ is the corresponding exact scattering eigenstate of the coupled system. $V$ is the coupling Hamiltonian between the nanostructure and the ideal 1D leads. A methodology for numerically solving Lippmann-Schwinger equations such Eq. (\ref{lippmann}) within a tight-binding framework  
is described in Appendix A of Ref. \onlinecite{Azari}.

In the present work the ideal leads are represented by 1D tight binding chains. Each site of each chain is assumed to have 6 orbitals, including spin. The on-site energies of these chain orbitals are the same as the corresponding atomic orbital energies $H^{i}_{\alpha \alpha}$ of the $ 6p_x,6p_y$ and $6p_z$ orbitals of the Bi atoms described by the Hamiltonian Eq. (\ref{Ham}). 
Only nearest neighbor Hamiltonian matrix elements {\em between like orbitals} of the 1D chains and between the chains and adjacent Bi atoms of the nanostructure are assumed to be non-zero. For simplicity, all of these nearest neighbor Hamiltonian matrix elements are assumed to have the same value $t= -2.0$eV. 

\section{Results}
\label{Results}
\subsection{Predictions of model of Section \ref{Model} for ideal Bi$_{105}$ dome}
\begin{figure}[t]
\centering
\includegraphics[width=0.7\linewidth]{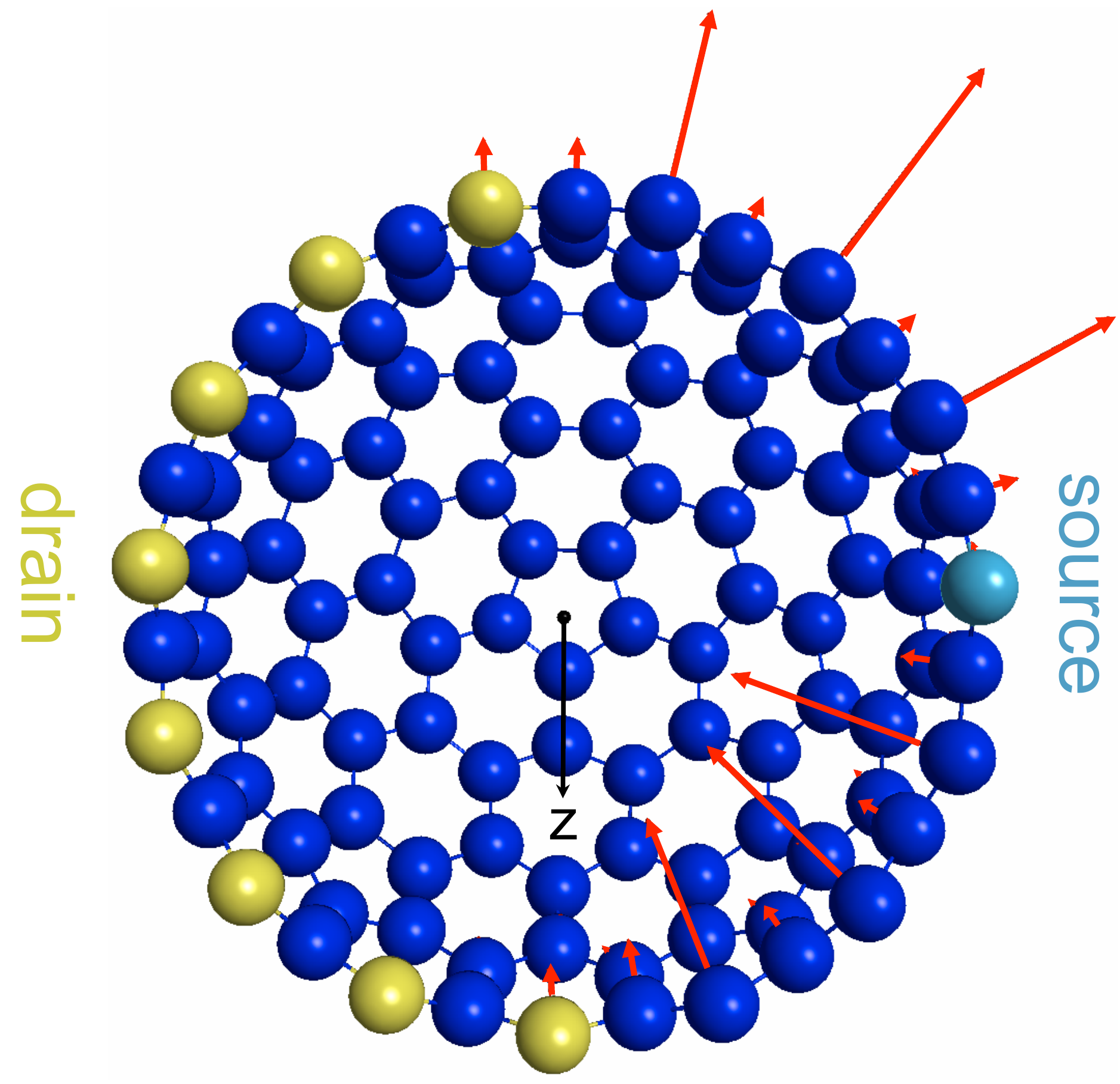}
\caption{(Color online).
View of the concave side of an ideal 105 atom bismuthene dome with zigzag edge 
formed by truncating a 180 atom 
spherical bismuthene fullerene without subsequent relaxation.
Light blue (chartreuse) atoms are edge atoms connected directly to the electron source (drain). 
Dark blue atoms are not contacted directly by either the source or drain.
The electrons entering the device from the source electrode are initially 
spin-{\em un}polarized.
Red arrows indicate the relative magnitudes and directions of the electron spin polarizations
induced on the individual atoms by an electron flux through the structure
from the electron source to the drain if the Fermi energy is 0.4 eV above
the valence band edge of bulk bismuthene on SiC. The black arrow is the z-axis.
Image prepared using Macmolplt software.\cite{MacMolPlt}}
\label{idealhemi} 
\end{figure}

\begin{figure}[t]
\centering
\includegraphics[width=1.0\linewidth]{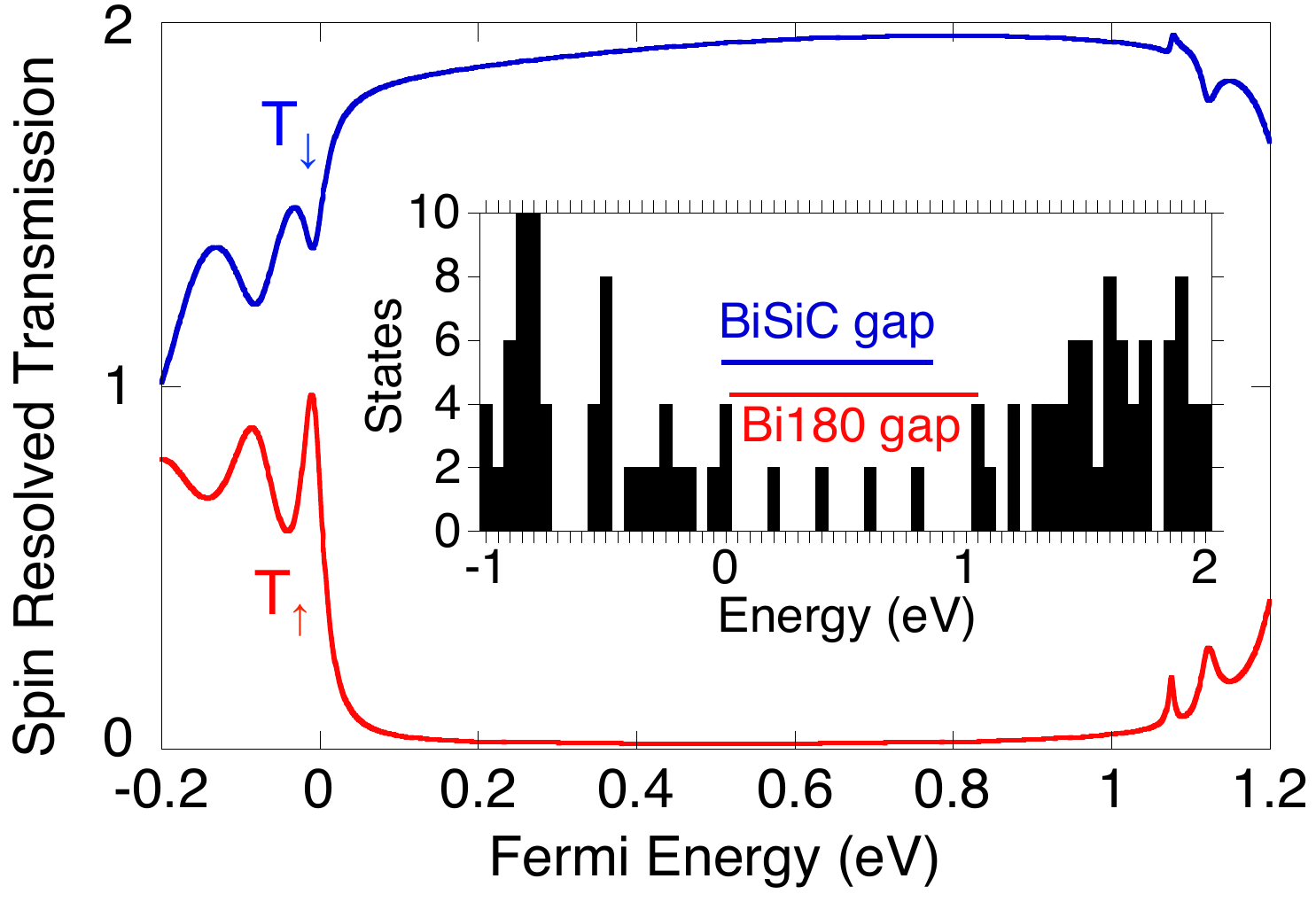}
\caption{(Color online).
Spin-resolved Landauer transmission probabilities $T_{\uparrow}$ and 
$T_{\downarrow}$ of spin up and spin down electrons
exiting through the drain contact of the device shown in Fig. \ref{idealhemi} at the Fermi energy assuming that spin unpolarized
electrons enter from the electron source. Inset: Histogram of eigenstates of the Hamiltonian of the device shown in Fig. \ref{idealhemi}
when it is decoupled from the leads. The band gap predicted by the present model for planar bismuthene on SiC is shown in blue.
The spectral gap predicted by the model for the spherical 180 atom bismuthene fullerene geometry is shown in red. The states in the
spectral gap are edge states.}
\label{Transidealhemi} 
\end{figure}

The predictions of the tight binding model  described in Section \ref{Model} for a curved topological insulator
whose structure is that of an ideal 105 atom bismuthene fullerene dome formed by truncating a 180 atom
bismuthene sphere without any subsequent relaxation of the atomic geometry
are shown in Figs. \ref{idealhemi} and \ref{Transidealhemi}. The geometry considered is shown in
Fig.  \ref{idealhemi} where the electron source (drain) electrode is attached to the light blue 
(chartreuse) colored bismuth edge atoms.

The calculated  electronic spectrum of this structure when the bismuthene is decoupled
from the contacts is shown in the inset of Fig. \ref{Transidealhemi}. The states in the spectral gap
of the 180 atom
bismuthene fullerene (marked in red) are edge states that are confined to the vicinity of the edge
of the 105 atom bismuthene fullerene dome. 

As has been discussed by Sheng {\em et al}.,\cite{Sheng2005} it is possible for a strong Rashba coupling to close the topological gap in some systems. However, it has been well established both theoretically and experimentally that the Rashba term, although large, is not sufficient to close the topological gap in planar bismuthene on SiC.\cite{Hsu2015,Reis2017,Dominguez2018,GLi2018,GK2018,Canonico2019,Azari2019,Hao2019,Stuhler2020}
Since the Rashba effect has a similar origin (the Bi-Si interaction) in our model system as in the planar bismuthene on SiC, it is reasonable to expect the Rashba effect not to close the topological gap in the present system either.

When contacts are attached to the 105 atom bismuthene fullerene dome as shown in Fig. \ref{idealhemi} and an electric
current flows, the calculated electron spin polarizations induced on the Bi atoms 
by the electric current are shown by red arrows in Fig. \ref{idealhemi}. Here the Fermi energy 
is at 0.4eV in the inset of Fig. \ref{Transidealhemi}, i.e., within the spectral gap of the complete 180 atom spherical
bismuthene fullerene. The electrons entering the device from the source electrode are initially 
spin-{\em un}polarized. However, as is seen in Fig. \ref{idealhemi}, strong spin polarizations are induced by the electric current on the 
bismuth atoms at the edges of the 105 Bi atom dome between the source and drain electrodes, as is expected for the spin polarization induced by electric currents
carried by edge states in quantum spin Hall systems. 

Notice that in Fig.  \ref{idealhemi} the spin polarizations point radially
outward from the center of the dome for electrons traveling counterclockwise along the edge
of the dome from source to drain but radially inward for electrons traveling clockwise along the edge.
Because of this, for the arrangement of electrodes shown in Fig.  \ref{idealhemi}, the spin polarizations
of {\em both} the clockwise and counterclockwise-moving edge states point in approximately the {\em same} direction
(i.e.,the negative $z$-direction) where these two edges feed into the drain electrode. In other words, {\em both} of the
edge states feed spin down electrons into the drain contact. Thus all of the electrons entering the drain electrode
have approximately spin down polarization. Consequently, this {\em two}-terminal curved topological insulator device functions
as a nearly perfect spin filter. 

This can be seen quantitatively in Fig. \ref{Transidealhemi} where for Fermi energies throughout most of the spectral gap of
the 180 atom bismuthene fullerene the spin down transmission  
$T_{\downarrow}$ of the 105 bismuth atom dome is close to 2 while the spin up transmission $T_{\uparrow}$ is close to zero.
The maximum value of spin polarization of the electrons entering the drain 
(defined as $T_{\downarrow}/(T_{\downarrow}+T_{\uparrow})$) is $\sim 0.99$ in Fig. \ref{Transidealhemi}. [More sophisticated measures of the spin polarization are in general possible, as described, for example, by Nikoli\'{c} et al.\cite{Nikolic2005} However, in this paper our objective is to show that the curved topological insulators can function as nearly perfect spin filters. For this purpose it is sufficient to demonstrate that $T_{\downarrow}/(T_{\downarrow}+T_{\uparrow})$ is close to 1 (in some range of Fermi energy values) since this implies that the z-component of the spin polarization of electrons entering the drain is much larger than the other components. Our calculations accomplish this.]

By contrast, a {\em planar} 2D {\em two}-terminal bismuthene on SiC topological insulator device does not exhibit such spin filtering
in a two-terminal arrangement because for it edge states traveling on opposite edges always have opposite spin polarizations and
the directions of those spin polarizations can not become aligned.  

The structure of the bismuthene dome considered in Figs.  \ref{idealhemi} and \ref{Transidealhemi} is idealized,
being an {\em unrelaxed} truncated portion of the 180 atom bismuthene fullerene sphere. Also the tight binding model used
(that described in Section \ref{Model}) is a generalization of the tight-binding models of {\em infinite} 2D bismuthene on SiC;
no adjustment of the tight binding parameters due to the presence of the edge is 
included in that model. 

\subsection{Predictions of improved model for Bi$_{105}$Si$_{105}$H$_{15}$ dome }

We now consider the more realistic Bi$_{105}$Si$_{105}$H$_{15}$ dome nanostructure shown in Fig. \ref{hemi}, taking into account 
the relaxed  geometry as well as the influence of the Si and H on the 
electronic structure of the bismuthene. We also provide an improved treatment of the Bi$_{105}$Si$_{105}$H$_{15}$ dome's edge electronic structure.  We accomplish this by modifying the tight binding model described in Section \ref{Model} so as to bring the bismuthene density of states (DOS) that the model predicts near the Fermi level (in the absence of the spin-orbit and Rashba terms of the Hamiltonian, Eq. \ref{Ham}) into agreement with that predicted by our density functional theory-based calculations for the relaxed Bi$_{105}$Si$_{105}$H$_{15}$ dome nanostructure.
\begin{figure}[t]
\centering
\includegraphics[width=1.0\linewidth]{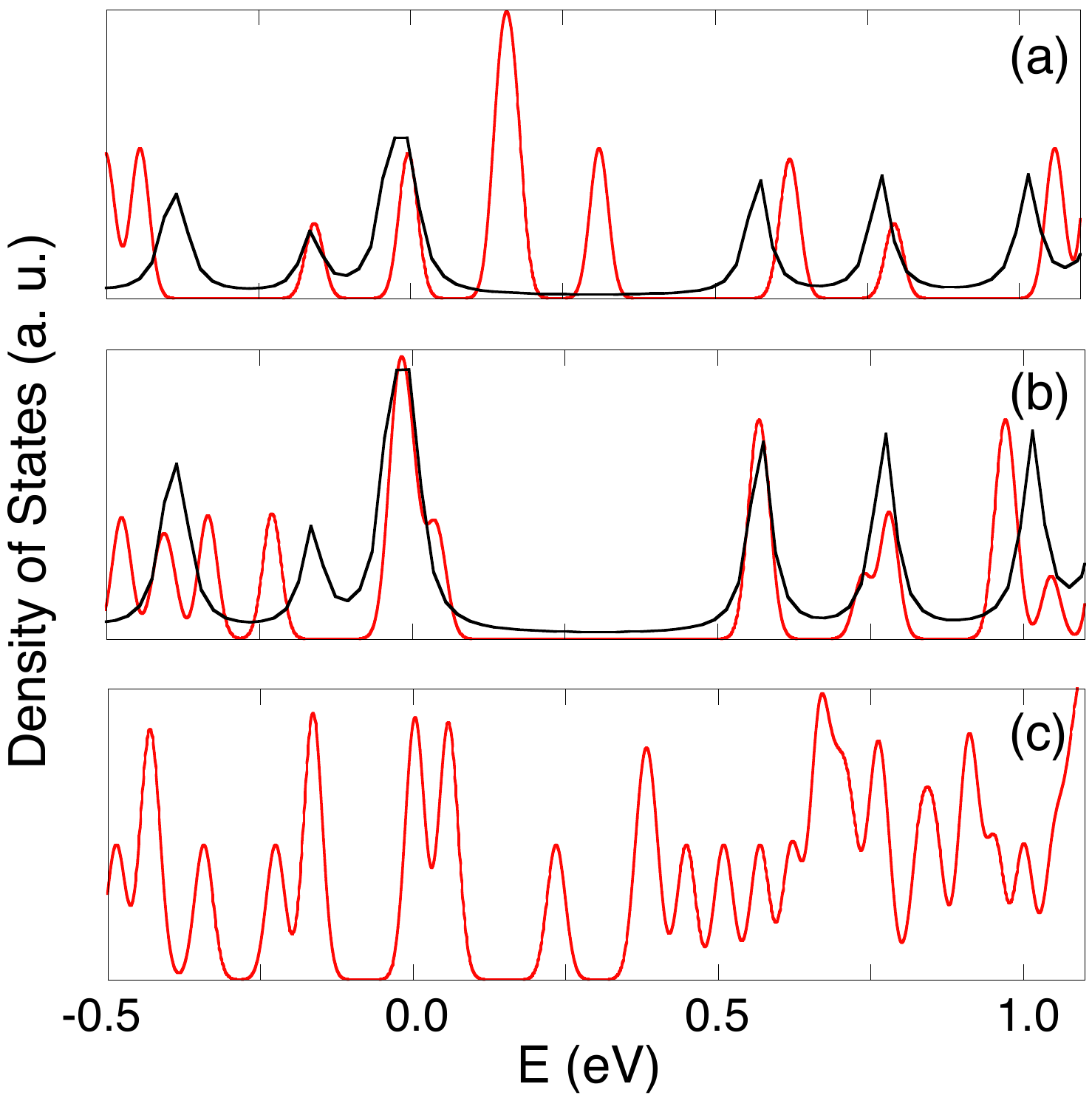}
\caption{(Color online). Calculated bismuth density of states vs. energy measured from the Fermi level ($E=0$) for the relaxed Bi$_{105}$Si$_{105}$H$_{15}$ dome shown in Fig.\ref{hemi}. (a): Black curve: Prediction of density functional theory
(B3PW91 functional with Lanl2DZ effective core potential and basis sets). Red curve: Prediction of the tight-binding model described in Section \ref{Model} 
 omitting the spin-orbit and Rashba terms from the Hamiltonian, Eq. \ref{Ham}. (b): Black curve: As in (a). Red curve: Prediction of the tight-binding model described in Section \ref{Model} omitting the spin-orbit and Rashba terms from the Hamiltonian, Eq. \ref{Ham}, but with the onsite energies of all of the orbitals of the 45 Bi atoms closest to the edge of the structure upshifted by 1.4 eV. (c) Red curve: As in (b) but including the spin-orbit and Rashba terms in the Hamiltonian.}
\label{DOS} 
\end{figure}
\begin{figure}[b]
\centering
\includegraphics[width=1.0\linewidth]{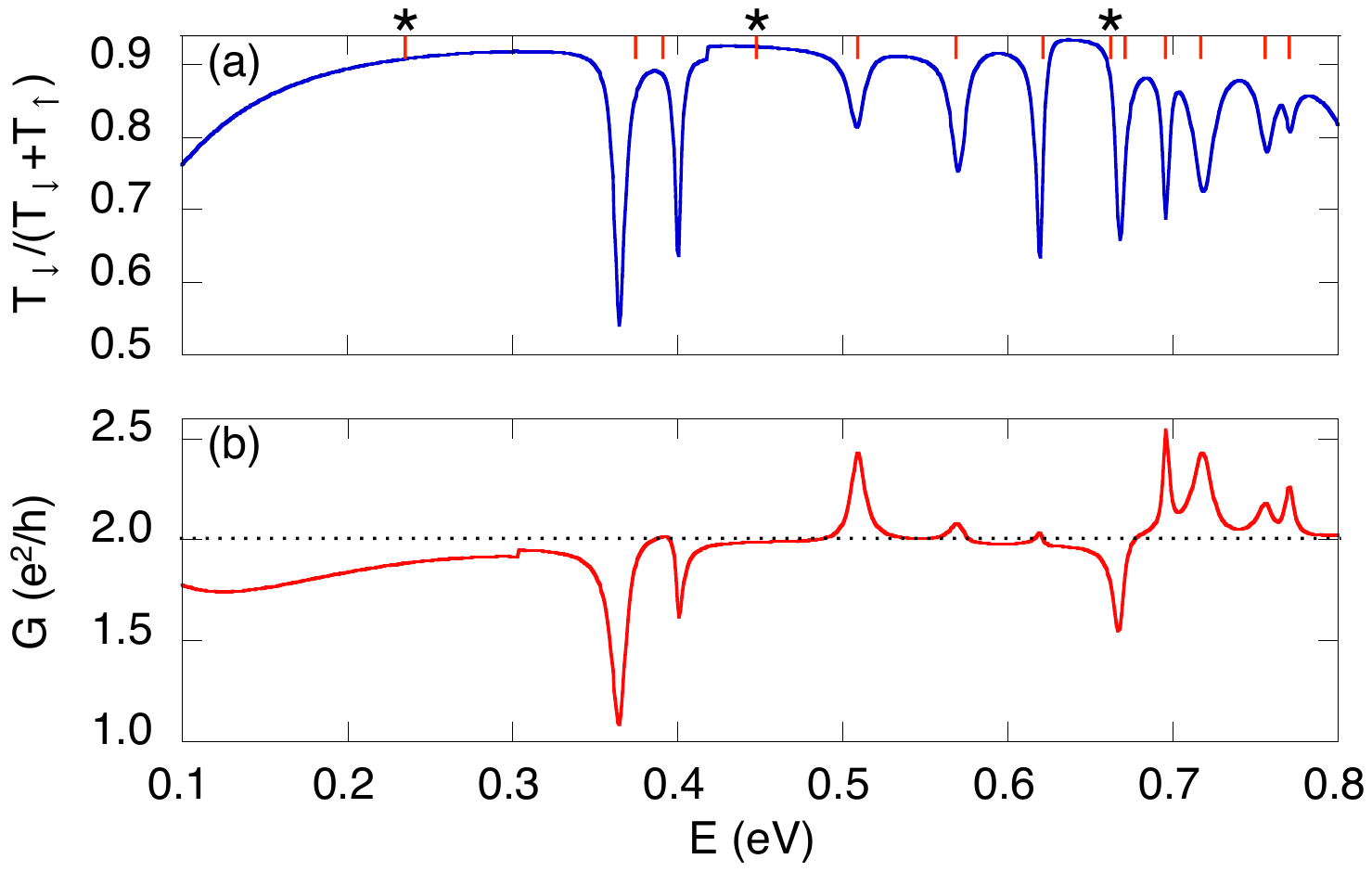}
\caption{(Color online). (a): Blue curve: Calculated spin polarization $T_{\downarrow}/(T_{\downarrow}+T_{\uparrow})$) of electrons entering the electron drain electrode from the Bi$_{105}$Si$_{105}$H$_{15}$ dome vs. Fermi energy, assuming that spin-unpolarized electrons enter the Bi$_{105}$Si$_{105}$H$_{15}$ dome from the source. The source and drain electrodes are connected to the bismuth atoms of the dome as in Fig. \ref{idealhemi}. The model Hamiltonian is as for Fig. \ref {DOS}(c). The energies of the eigenstates of the Hamiltonian when the Bi$_{105}$Si$_{105}$H$_{15}$ dome is disconnected from the leads are shown by the red tick marks at the top of the figure. The edge states are indicated by asterisks ($\star$). (b) Red curve: Calculated conductance $G = (T_{\downarrow}+T_{\uparrow})e^2/h$ of the Bi$_{105}$Si$_{105}$H$_{15}$ dome vs. Fermi energy.
}
\label{filter} 
\end{figure}

The black curve in Fig.\ref{DOS}(a) shows the bismuthene DOS for the Bi$_{105}$Si$_{105}$H$_{15}$ dome computed within DFT while the red curve shows the corresponding DOS predicted by the model presented in  Section \ref{Model} 
 omitting the spin-orbit and Rashba terms from the Hamiltonian, Eq. \ref{Ham}.  A striking difference between these two curves is the presence of the strong peaks in the red curve near $E = 0.15$ and 0.3eV, within the prominent gap in the DFT DOS (black curve) that extends from the Fermi level ($E=0$) to $E \sim 0.57$ eV. The other peaks of the red curve in Fig.\ref{DOS}(a) are in remarkably good agreement with the peaks of the black curve, especially in view of the fact that no attempt was made to fit the model  in  Section \ref{Model} to DFT calculations for the Bi$_{105}$Si$_{105}$H$_{15}$ dome. The anomalous peaks of the red curve in Fig.\ref{DOS}(a) (those near $E = 0.15$ and 0.3eV) are due to edge states. We note in passing that the existence of the edge states near $E = 0.15$ and 0.3eV does not imply that the system supporting them is a topological insulator {\em in the absence of spin-orbit coupling} since in the absence of spin-orbit and Rashba terms it does not exhibit a spin Hall effect.

The presence of the anomalous edge state peaks of the red curve in Fig.\ref{DOS}(a) (near $E = 0.15$ and 0.3eV) suggests that the model presented in  Section \ref{Model} does not describe the electronic structure of the edge of Bi$_{105}$Si$_{105}$H$_{15}$ dome adequately. We find that the DOS predicted by the tight binding model can be brought into good agreement with that of DFT if we incorporate into the model an effective edge potential such that the energies $H^{i}_{\alpha,\alpha}$ of all of the atomic orbitals of the 45 Bi atoms that are closest to the edge of the Bi$_{105}$Si$_{105}$H$_{15}$ dome are shifted upwards in energy by 1.4eV. The tight-binding model modified in this way (still omitting the spin-orbit and Rashba terms) yields the DOS plotted in red in Fig.\ref{DOS}(b). As can be seen in Fig.\ref{DOS}(b) a gap has opened in the DOS predicted by the model, matching that predicted by DFT (the black curve in Fig.\ref{DOS}(b)) and reasonably good agreement between the other low energy features of the DOS predicted by DFT and the modified model is retained. Interestingly, varying the values of the other parameters of the model in Section \ref{Model} and/or changing the profile of the edge potential from that described above does not yield significantly better agreement between the DOS predicted by the modified model and that yielded by DFT than that displayed in Fig.\ref{DOS}(b)). 
 
Importantly, we conclude that near the edge an effective edge potential should be included in our tight-binding Hamiltonian whereas the tight-binding hopping parameters do not require substantial modification.
The physical reasons why the hopping parameters do not require adjustment appear to be
(i) Our tight binding hopping parameters include the $p_z$ Bi valence orbitals in addition to the $p_x$ and $p_y$ orbitals  that are sufficient to describe planar bismuthene on SiC. 
(ii) The Slater-Koster type dependence of the hopping parameters on the Bi-Bi bond orientation implemented in Table II appears to be adequate for this system, and 
(iii) The onsite Hamiltonian as parameterized in Table I succeeds in capturing appropriately the Bi $p$-orbital level splitting induced by the coupling between the neighboring Bi and Si atoms even for the curved bismuthene geometry.
 
Reintroducing the spin-orbit and Rashba terms of Eq.\ref{Ham} into the modified tight binding model yields the DOS shown in red in Fig.\ref{DOS}(c). That the spin-orbit and Rashba terms have resulted in the presence of electronic states in the energy range of the spectral gap of Fig.\ref{DOS}(b) is not surprising since the magnitude of the spin orbit coupling parameter $\zeta_1 = 1.8$eV in Eq. \ref{intraSO} substantially exceeds the size of the spectral gap in Fig.\ref{DOS}(b). Never the less we regard the modification of the tight-binding model prompted by our comparison between the DOS of DFT and the model in the absence of spin orbit coupling as a significant improvement of the model as applied to the Bi$_{105}$Si$_{105}$H$_{15}$ dome and especially the properties of its edge. This becomes apparent when one considers spin filtering by the Bi$_{105}$Si$_{105}$H$_{15}$ dome, as will be explained next.

 In Fig.\ref{filter}(a) the blue curve shows the calculated spin polarization of electrons injected into the drain electrode from the Bi$_{105}$Si$_{105}$H$_{15}$ dome at the Fermi energy $E$ in the linear response regime, assuming that the electrons entering the dome from the source electrode are spin unpolarized.
The source and drain electrodes are connected to the bismuth atoms of the dome as in Fig. \ref{idealhemi}. The energies of the Hamiltonian eigenstates of the dome when it is disconnected from the leads are shown by the red ticks at the top of Fig.\ref{filter}(a). The eigenstates that are edge states are marked by asterisks ($\star$). Values of the spin polarization as high as $T_{\downarrow}/(T_{\downarrow}+T_{\uparrow})\sim 0.93$ can be seen in Fig.\ref{filter}(a). The mechanism responsible for the largest values of the spin polarization is similar to that illustrated in Fig.\ref{idealhemi}, i.e., 
the spin polarizations associated with the electric currents carried by the edge states point radially
outward from the the dome for electrons traveling counterclockwise along the edge
of the dome from source to drain but radially inward for electrons traveling clockwise along the edge.
Because of this, the spin polarizations
of the clockwise and counterclockwise-moving edge become approximately aligned
where these two edges feed into the drain electrode and the dome functions
as a nearly perfect spin filter. The eigenstates of the Hamiltonian that are not edge states (indicated in Fig.\ref{filter}(a) by red ticks without 
asterisks) do not support this spin filtering mechanism. Thus when the Fermi level is close in energy to any of these non-edge states the value 
of the spin polarization of the electrons entering the drain is sharply reduced. Notice that unlike in Fig.\ref{Transidealhemi} where all of the states in the bulk spectral gap are edge states, in Fig.\ref{filter}(a) the edge states are interspersed with non-edge states that are in fact the majority. However, since the non-edge states have very small amplitudes on the bismuth atoms at the edge of the dome, they couple very weakly to the electrodes and consequently the transport resonances associated with them are narrow. Because of this the influence of the edge states and the strong spin filtering associated with them predominates at energies between the spin polarization dips associated with the non-edge states even at energies between non-edge states where there is no edge eigenstate. 

This is further clarified by considering the calculated conductance $G = (T_{\downarrow}+T_{\uparrow})e^2/h$ of the Bi$_{105}$Si$_{105}$H$_{15}$ dome vs. Fermi energy $E$ shown by the red curve in Fig.\ref{filter}(b). This plot shows sharp conductance resonances or antiresonances at the energies of the non-edge Hamiltonian eigenstates. However, no such features in the conductance are visible at the energies of the edge eigenstates. This is due to the very strong coupling of the edge eigenstates to the leads that results in very strong broadening of the edge eigenstates. This in turn means that the edge eigenstates influence the spin transport strongly over wide energy ranges that are interrupted by narrow regions where the effects of the non-edge eigenstates are crucial.

\section{Summary}
\label{Summary}

We have investigated theoretically the properties of curved nanostructures derived from 2D topological insulators, a topic that has 
not previously received experimental or theoretical attention. We have shown that curved geometries make it possible to manipulate the spin polarizations of electron edge states in ways that are not possible for planar systems, opening the way for the realization of previously unanticipated devices. In particular, we have shown that it is possible to bring the spin polarizations of electron edge states traveling along opposite edges of curved nanostructures into alignment and thus to realize nearly perfect two-terminal spin filters operating in the absence of magnetic fields. 

Our study has combined the construction and application of appropriate tight-binding models and
density functional theory-based calculations. We generalized our previous tight binding model of the wide band gap planar topological insulator bismuthene
on SiC to curved geometries. Our transport calculations based on the resulting tight binding model applied to an ideal spherical bismuthene dome with a free zigzag edge showed this model system to exhibit quantum spin Hall physics: We showed this model to support edge states propagating along the edge of the dome within the bulk band gap of planar bismuthene on SiC. These edge states have radially oriented spin polarizations that point towards or away from the center of the dome, depending on whether the edge state travels in the clockwise or counter-clockwise direction. Thus, unlike for planar 2D topological insulators (for which the directions of the spin polarizations of the edge states are fixed), the directions of spin polarizations of the edge states of the curved structure can vary, controlled by the nanostructure's geometry. Because of this it is possible for the spin polarizations of the clockwise and counter-clockwise edge states traveling from the electron source electrode to the drain to be parallel where the edge states enter the drain. Thus we predict that nearly perfect two terminal spin filters operating in the absence of magnetic fields can, in principle, be constructed from curved topological insulators. 

Our density functional theory-based calculations have addressed the question whether chemically modified bismuthene domes exhibiting such spintronic properties can be stable. We find that a bismuthene dome of 105 bismuth atoms with a zigzag edge can be stabilized by the adsorption of a silicon atom to each bismuth atom on the concave surface of the dome if the edge silicon atoms are passivated with hydrogen. We have carried out density functional theory based calculations of the bismuth density of states of this Bi$_{105}$Si$_{105}$H$_{15}$ structure and have modified our tight binding model so that its predicted density of states approximately matches the corresponding density functional theory based result. We find that the resulting modified tight binding model of Bi$_{105}$Si$_{105}$H$_{15}$ exhibits both edge states and non-edge states at energies in the bulk band gap of planar bismuthene on SiC. However, our transport calculations for this system predict the influence of the edge states and the strong spin filtering mechanism associated with them to predominate for several ranges of the Fermi energy, including some Fermi energy ranges lying between the energies of consecutive non-edge states.

The results of our investigation show that curved nanostructures based on 2D topological insulators should have spintronic properties that differ qualitatively from those of planar 2D topological insulators and that these distinctive properties may have important practical applications. We have identified a nanostructure that we predict to be stable and to exhibit these properties. Based on our work, it is clear that experimental and theoretical studies exploring this topic further are warranted and would be of considerable interest.
 
\begin{acknowledgments}
This research was supported by NSERC, Westgrid,
and Compute Canada.
\end{acknowledgments}
%% ----------------------------------------------------------------------------------------------------
{

\end{document}
\begin{thebibliography}{333}
{\footnotesize 

\bibitem{KM1}Kane, C. L., and E. J. Mele, Quantum Spin Hall Effect in Graphene, 
Phys. Rev. Lett. {\bf 95}, 226801 (2005).
\bibitem{KM2}Kane, C. L., and E. J. Mele, Z$_2$ Topological Order and the Quantum 
Spin Hall Effect, Phys. Rev. Lett. {\bf 95}, 146802 (2005).
\bibitem{Bernevig2006}B. A. Bernevig, T. L. Hughes, S.-C. Zhang, 
Quantum Spin Hall Effect and Topological Phase Transition in HgTe Quantum Wells,
Science 314, 1757 (2006).
\bibitem{FuKane2007}L. Fu, C. L. Kane, Topological insulators with inversion symmetry,
Phys. Rev. B {\bf 76}, 045302 (2007).
\bibitem{Konig2007}M. K\"{o}nig, S. Wiedmann, C. Br\"{u}ne, A. Roth, H. Buhmann, L.
W. Molenkamp, X. L. Qi, S. C. Zhang,
Quantum Spin Hall Insulator State in HgTe Quantum Wells,
Science  {\bf 318}, 766 (2007).
\bibitem{Moore2007}J. E. Moore and L. Balents, Topological invariants of time-reversal-invariant band structures,
Phys. Rev. B {\bf 75}, 121306(R) (2007).
\bibitem{Hasan2010}M. Z. Hasan, C. L. Kane, Colloquium: Topological insulators, 
Rev. Mod. Phys. {\bf 82}, 3045 (2010). 
\bibitem{Qi2011}X.-L. Qi, S.-C. Zhang, Topological insulators and superconductors,
Rev. Mod. Phys. {\bf 83}, 1057 (2011).
\bibitem{Ando2013}Y. Ando, Topological Insulator Materials,
J. Phys. Soc. Jpn. {\bf 82}, 102001 (2013)
\bibitem{TIcourse}J. K. Asb\'{o}th, L. Oroszl\'{a}ny, A. P\'{a}lyi,
Sec. 8.4, {\em A Short Course on Topological Insulators}, 
vol. 919, {\em Lecture Notes in Physics}, 
Springer, 2016. DOI 10.1007/978-3-319-25607-8

\bibitem{Hsu2015}C.-H. Hsu, Z.-Q. Huang, F.-C. Chuang, C.-C. Kuo, Y.-T. Liu, H. Lin, A. Bansil,
The nontrivial electronic structure of Bi/Sb honeycombs on SiC(0001),
New J. Phys. {\bf 17}, 025005 (2015).

\bibitem{Reis2017}F. Reis, G. Li, L. Dudy, M. Bauernfeind, S. Glass, W. Hanke, R. Thomale,
J. Sch\"{a}fer, R. Claessen, Bismuthene on a SiC substrate:
A candidate for a high-temperature
quantum spin Hall material, Science {\bf 357}, 287 (2017). 

\bibitem{Dominguez2018}F. Dominguez, B. Scharf, G. Li, J. Sch\"{a}fer, R. Claessen, W. Hanke, R. Thomale, E. M. Hankiewicz,
Testing Topological Protection of Edge States in Hexagonal Quantum Spin Hall Candidate Materials,
Phys. Rev. B {\bf 98}, 161407(R) (2018).

\bibitem{GLi2018}G. Li, W. Hanke, E. M. Hankiewicz, F. Reis, J. Sch\"{a}fer, R. Claessen, C. Wu, R. Thomale,
Theoretical paradigm for the quantum spin Hall effect at high temperatures,
Phys. Rev. B {\bf 98}, 165146 (2018).

\bibitem{GK2018}G. Kirczenow, Perfect and imperfect conductance quantization and transport resonances of 
two-dimensional topological-insulator quantum dots with normal conducting leads and contacts,
Phys. Rev. B {\bf 98}, 165430 (2018).

\bibitem{Canonico2019}L. M. Canonico, T. G. Rappoport, and R. B. Muniz,
Spin and Charge Transport of Multiorbital Quantum Spin Hall Insulators,
Phys. Rev. Lett. {\bf 122}, 196601 (2019).

\bibitem{Azari2019}M. Azari and G. Kirczenow, Valley polarization reversal and spin ferromagnetism and antiferromagnetism in quantum dots of the topological insulator monolayer bismuthene on SiC
Phys. Rev. B {\bf 100}, 165417 (2019).

\bibitem{Hao2019}X. Hao, F. Luo, S. Zhai, Q. Meng, J. Wu, L. Zhang, T. Li, Y. Jia, M. Zhou,
Strain-engineered electronic and topological properties of bismuthene on SiC(0001) substrate,
Nano Futures {\bf 3}, 045002 (2019).

\bibitem{Stuhler2020}
R. St\"{u}hler, F. Reis, T. M\"{u}ller, T. Helbig, T. Schwemmer, R. Thomale, J. Sch\"{a}fer, R. Claessen,
Tomonaga-Luttinger liquid in the edge channels of a quantum spin Hall insulator, 
Nature Physics {\bf 16}, 47 (2020).

\bibitem{Su2002}C. Su, H.-T. Liu and J.-M. Li,
Bismuth nanotubes: potential semiconducting nanomaterials,
Nanotechnology {\bf 13}, 746 (2002).

\bibitem{Zdetsis2010}A. D. Zdetsis,
Theoretical Predictions of a New Family of Stable Bismuth and Other Group 15 Fullerenes,
J. Phys. Chem. C {\bf 114}, 10775 (2010).

\bibitem{Kharissova2012}O. V. Kharissova, M. Osorio,
M. S. V\'{a}zquez, B. I. Kharisov,
Computational chemistry calculations of stability
for bismuth nanotubes, fullerene-like structures
and hydrogen-containing nanostructures,
J. Mol. Model. {\bf 18}, 3981 (2012).

\bibitem{Jin2017}K.-H. Jin, S.-H. Jhi, F. Liu,
Nanostructured topological state in bismuth nanotube arrays: 
inverting bonding-antibonding levels of molecular orbitals,
Nanoscale {\bf 9}, 16638 (2017).

\bibitem{Li2001}Y. Li, J. Wang, Z. Deng, Y. Wu, X. Sun, D. Yu and P. Yang,
Bismuth nanotubes: A rational low-temperature synthetic route,
J. Am. Chem. Soc. {\bf 123}, 9904 (2001).

\bibitem{Yang2003}B. Yang, C. Li, H. Hu, X. Yang, Q. Li and Y. Qian,
A room-temperature route to bismuth nanotube arrays,
Eur. J. Inorg. Chem. {\bf 2003}, 3699 (2003).


\bibitem{Li2006}L. Li, Y. W. Yang, X. H. Huang, G. H. Li, R. Ang, and L. D. Zhang, 
Fabrication and electronic transport properties of Bi nanotube arrays,
Appl. Phys. Lett. {\bf 88}, 103119 (2006). 

\bibitem{Yang2008}D. Yang, G. Meng, Q. Xu, F. Han, M. Kong, and L. Zhang,
lectronic Transport Behavior of Bismuth Nanotubes with a Predesigned Wall Thickness,
J. Phys. Chem. C {\bf 112}, 8614 (2008).

\bibitem{Ma2010}D. Ma, J. Zhao, Y. Li, X. Su, S. Hou, Y. Zhao, X. Hao, L. Li,
Organic molecule directed synthesis of bismuth nanostructures 
with varied shapes in aqueous solution and their optical characterization,
Colloids Surf. A {\bf 368}, 105 (2010).

\bibitem{Derrouiche2010a}S. Derrouiche, C. Z. Loebick, and L. Pfefferle, 
Optimization of Routes for the Synthesis of Bismuth Nanotubes: Implications for
Nanostructure Form and Selectivity,
J. Phys. Chem. C {\bf 114}, 3431 (2010).

\bibitem{Derrouiche2010b}S. Derrouiche, C. Z. Loebick, C. Wang, and L. Pfefferle,
Energy-Induced Morphology Changes in Bismuth Nanotubes,
J. Phys. Chem. C {\bf 114}, 4336 (2010).

\bibitem{Boldt2010}
R. Boldt, M. Kaiser, D. K\"{o}hler, F. Krumeich and M. Ruck,
High-Yield Synthesis and Structure of Double-Walled Bismuth-Nanotubes,
Nano Lett. {\bf 10}, 208 (2010).

\bibitem{Frisch}
M. J. Frisch, G. W. Trucks, H. B. Schlegel, G. E. Scuseria, M. A. Robb, J. R. Cheeseman, G. Scalmani, V. Barone, G. A. Petersson {\em et al.}, the GAUSSIAN 16 Revision: A.03 computer code was used.

\bibitem{MacMolPlt}
B. M. Bode, M. S. Gordon, Macmolplt: a graphical user interface for GAMESS, 
J. Mol. Graphics and Modeling
{\bf 16}, 133 (1998).

\bibitem{SK1954}J. C. Slater and G. F. Koster, Simplified LCAO Method for the Periodic Potential Problem,
Phys. Rev. {\bf 94}, 1498 (1954).
\bibitem{Ammeter}
{J. H. Ammeter, H. B. Buergi, J. C. Thibeault, and R. Hoffmann,
J. Am. Chem. Soc. {\bf 100}, 3686 (1978).}

\bibitem{PRBrapid} F. Rostamzadeh Renani and G. Kirczenow, Ligand-Based Transport Resonances of Single-Molecule-Magnet Spin Filters: Suppression of Coulomb Blockade and Determination of Easy-Axis Orientation, Phys. Rev. B {\bf 84}, 180408(R) (2011).

\bibitem{PRB} F. Rostamzadeh Renani and G. Kirczenow, Tight Binding Model of Mn12 Single Molecule Magnets: Electronic and
Magnetic Structure and Transport Properties, Phys. Rev. B {\bf 85}, 245415 (2012).

\bibitem{CS}E. U. Condon and G. H. Shortley, {\em The Theory of Atomic Spectra}, Cambridge University Press, London, 1935.

\bibitem{BR1} Y. A. Bychkov and E. I. Rashba, Properties of a 2D electron gas with lifted spectral 
degeneracy, Pis'ma Zh. Eksp. Teor. Fiz. {\bf 39}, 66 (1984); JETP Lett. {\bf 39}, 78 (1984).

\bibitem{BR2} Y. A. Bychkov and E. I. Rashba, Oscillatory effects and the magnetic susceptibility 
of carriers in inversion layers, J. Phys. C {\bf 17}, 6039 (1984).

\bibitem{Kittel}C. Kittel, {\em Quantum Theory of Solids}, Wiley, New York, 1963, p. 181.

\bibitem{Econ81}E. N. Economou and C. M. Soukoulis, Static Conductance and Scaling Theory of Localization in One Dimension,
Phys. Rev. Lett. {\bf 46}, 618 (1981).

\bibitem{Fish81}D. S. Fisher and P. A. Lee, Relation between conductivity and transmission matrix,
Phys. Rev. B{\bf 23}, 6851 (1981).

\bibitem{Azari}M. Azari and G. Kirczenow, Gate-tunable valley currents, non-local resistances 
and valley accumulation in bilayer graphene nanostructures, Phys. Rev. B{\bf 95}, 195424 (2017).

\bibitem{Sheng2005}L. Sheng, D. N. Sheng, C. S. Ting, and F. D. M. Haldane,
Nondissipative Spin Hall Effect via Quantized Edge Transport,
Phys. Rev. Lett. {\bf 95}, 136602 (2005)

\bibitem{Nikolic2005}B. K. Nikoli\'{c} and S. Souma, Decoherence of transported spin in multichannel spin-orbit-coupled spintronic devices:
Scattering approach to spin-density matrix from the ballistic to the localized regime,
Phys. Rev. B{\bf 71}, 195328 (2005)


}
\end{thebibliography}
